\documentstyle[eqsecnum,aps,epsf,twocolumn]{revtex}
\begin{document}
\title{On the Ginzburg-Landau Free Energy of 
Charge Density Waves\\ with 
a Three-Dimensional Order}
\author{Masahiko Hayashi$^{1,3}$ and Hideo Yoshioka$^{2,3}$}
\address{$^{1}$Graduate School of 
Information Sciences, Tohoku University, 
Aramaki Aoba-ku, Sendai 980, Japan\\
$^{2}$Department of Physics, 
Kitauoyanishimachi, 
Nara Women's University, 
Nara 630-8506, Japan\\
$^{3}$Department of Applied Physics, 
Delft University of Technology,
2628CJ Delft, The Netherlands
}
\date{\today}
\wideabs{
\maketitle
\begin{abstract}
The effective free energy of 
a charge density wave (CDW) with a 
three-dimensional order is derived 
from a microscopic model 
(Fr\"olich model) 
based on the path integral method. 
Electron hoping and Coulomb interaction 
between chains are taken into account 
perturbatively, leading to an elastic 
interchain coupling of 
the CDW ordered state. 
\end{abstract}

%\pacs{Valid PACS appear here. PACS should always be input, }
}

\narrowtext

\section{Introduction}

After a pioneering study by Peierls \cite{Peierls}, 
charge density wave (CDW) has been 
a subject of continuous interest because of 
its unique nature as an electronic condensate. 
Especially, its behavior under an electric 
field was investigated by Fr\"olich \cite{Frolich} 
as a possible origin of superconductivity. 
Although it turned out that the CDW 
mechanism does not apply to the superconductivity, 
many interesting behaviors have been found in the 
electromagnetic response of 
CDW \cite{Gruner,Gorkov-Gruner}. 
Even today, experimental developments 
are adding intriguing new phenomena to the 
book of CDW. 
For example, AB effect in CDW's with columnar defects 
\cite{Latyshev}, field effect \cite{Adelman} and 
current effect \cite{Markovic} of CDW transport 
have been discovered 
although some theoretical developments are 
needed for the full understanding of these 
phenomena. 

In a preceding paper \cite{Hayashi-Yoshioka}, 
we have studied CDW's in a transverse (perpendicular to 
the chains) electric field and pointed out 
a possibility of 
a state similar to the mixed state of superconductors. 
Although interchain coupling 
plays an essential role in this theory, 
it was only phenomenologically 
introduced. 
The aim of the present paper is to give 
a microscopic foundation of the interchain coupling 
by deriving it from the electron-phonon model. 

Although there have been proposed several origins 
of the interchain coupling \cite{Horovitz,Saub}, 
the derivation of the Ginzburg-Landau (GL) type 
effective free energy taking them into account 
in a unified framework has not been given until now. 
We reexamine the effective free energy of the 
CDW taking account of the electron hopping 
and the Coulomb interaction between chains 
as sources of the interchain coupling. 
These effects are treated perturbatively 
employing the path integral formulation. 
It is shown that the free energy previously used by 
the present authors \cite{Hayashi-Yoshioka} can be 
derived from a microscopic model. 

For simplicity we limit our discussion to the 
case without impurity and commensurability pinning. 

\section{Free Energy of a CDW System}

We consider a three-dimensional (3-D) 
stack of 1-D conducting
chains, each of which is described by 
Fr\"olich model
(coupled electron-phonon system). 
At low temperatures, this system is
expected to show a 3-D CDW order due 
to interchain coupling \cite{Gruner,Gorkov-Gruner}.
We treat the Coulomb interaction 
and the electron hopping between chains 
as origins of interchain coupling. 
However the electron hopping is 
assumed to be small 
so that it does not break 
the one-dimensionality of the system \cite{hopping}. 

\subsection{One-dimensional electron-phonon system}
The action of the electron-phonon system 
$S_{\mbox{\scriptsize e-p}}$ is given by, 
\begin{eqnarray}
S_{\mbox{\scriptsize e-p}}&=
&\int_{0}^{\hbar \beta}d\tau \,\sum_{i} 
\biggl[
\nonumber\\
&&\int{\rm d}x\, 
\psi_{i\sigma}^{*}(x)\left(\hbar \partial_{\tau}
-\frac{\hbar^{2}}{2 m}\partial_{x}^{2}-\mu 
\right) \psi_{i\sigma}(x)\nonumber\\
&& +\sum_{q}b_{i}^{*}(q)\,\left(
\hbar \partial_{\tau} +\hbar \omega _{q}\right)
b_{i}(q)  \nonumber \\
&& +\frac{1}{\sqrt{L}}\sum_{k,q}g_{q}
a_{i\sigma }^{*}(k+q)a_{i\sigma
}(k)\left\{ b_{i}(q)+b_{i}^{*}(-q)\right\}\biggr],
\nonumber\\
\label{action-1}
\end{eqnarray}
where $i$ is the index numbering the chains, 
$L$ is the length of the chains 
and, $m$ and $-e$ are the mass 
and charge of an electron, respectively. 
The chemical potential $\mu$ is given by 
$\mu = \hbar^{2} k_{F}^{2}/(2 m)$ 
with $k_{F}$ being the Fermi wave number. 
We assume that $\mu$ is common to all the chains 
\cite{hopping}. 
The phonon frequency and electron-phonon 
coupling constant of wave number $q$ are denoted 
by $\omega _{q}$ and $g_{q}$, 
respectively. 
Electron field on the $i$-th chain with spin $\sigma$ 
is represented by 
$\psi_{i\sigma }(x,\tau)$ and its Fourier transformation is 
given by $a_{i\sigma}(k,\tau) = 
L^{-1/2}\int {\rm d}x\,\psi_{i\sigma }(x,\tau)\exp(-i k x)$. 
Note that the summation over $\sigma$ is suppressed 
in the present paper. 
The Fourier component of phonon field
of the $i$-th chain is denoted by $b_{i}(q)$ 
which is related to the displacement of ions $u_{i}(x)$ by, 
\begin{equation}
u_{i}(x)=\frac{i}{\sqrt{L}}\sum_{q}\alpha _{q}
\epsilon _{q}\left\{
b_{i}(q)+b_{i}^{*}(-q)\right\} {\rm e}^{iqx},  
\label{displacement}
\end{equation}
where $\alpha _{q}=\hbar /\sqrt{2\bar{\rho}_{M}
\hbar \omega _{q}}$ with 
${\ \bar{\rho}}_{M}$ being the average mass 
density of ions and $\epsilon_{q}=q/|q|$. 
Here we have considered only the 
phonons corresponding to the displacement of ions 
parallel to the chains. 

We consider the Peierls instability 
at the wave number $Q \equiv \pm 2k_{F}$ 
and treat only the phonons with 
wave number close to $\pm 2k_{F}$. 
It is convenient to
divide $\psi _{i\sigma}$ into two parts as 
$\psi _{i\sigma}={\rm e}^{ik_{F}x} R_{i\sigma}+
{\rm e}^{-ik_{F}x}L_{i\sigma}$, where $R_{i\sigma}$
and $L_{i\sigma}$ stand for right 
and left moving electrons, respectively, 
and are assumed to be slowly varying fields. 

The complex order parameter (or energy gap) 
of CDW is introduced by 
\begin{equation}
\Delta _{i}(x)=\frac{g}{\sqrt{L}}\sum_{q}\langle
b_{i}^{*}(-Q-q)+b_{i}(Q+q)\rangle {\rm e}^{iqx},  
\label{delta}
\end{equation}
where $\langle \cdots \rangle $ denotes the statistical average 
and we neglect the $q$-dependence of $g_{Q+q}$, 
writing it as $g$. 
Here we have assumed that the phonon field and 
the order parameter are classical fields 
which do not depend on imaginary time $\tau$. 
Then the action can be rewritten as, 
\begin{equation}
S_{\mbox{\scriptsize e-p}}=\int_{0}^{\hbar \beta }
d\tau \sum_{i}\int_{0}^{L}dx
\left[ \Psi_{i\sigma }^{\dagger }\cdot 
{\bf K}_{i}\cdot \Psi _{i\sigma }+
\frac{\hbar \omega_{Q}}{2g^{2}}|\Delta _{i}|^{2}\right] , 
\end{equation}
where 
\begin{eqnarray}
{\bf K}_{i}(x,\tau ) &=&\left( 
\begin{array}{cc}
\hat{k}_{i}^{(+)} & \Delta _{i}(x) \\ 
\Delta _{i}^{*}(x) & \hat{k}_{i}^{(-)}
\end{array}
\right) ,  \nonumber \\
\Psi _{i\sigma }(x,\tau ) &=&\left( 
\begin{array}{c}
R_{i\sigma }(x,\tau ) \\ 
L_{i\sigma }(x,\tau )
\end{array}
\right) , \\
\hat{k}_{i}^{(\pm )} &=&\hbar \partial_{\tau} 
-\frac{\hbar^{2}}{2m}\left( \pm ik_{F}+\partial_{x}
\right) ^{2}-\mu .
\end{eqnarray}
Here $q$-dependence of $\omega _{Q+q}$ 
is ignored as 
$\omega _{Q+q} \longrightarrow \omega _{Q}$. 
By applying the variation 
with respect to $\Delta _{i}^{*}(x)$  
we obtain self-consistent equation, 
\begin{equation}
\Delta _{i}(x) = - \frac{2 g^{2}}{\hbar \omega _{Q}}
\langle 
L_{i\sigma }^{*}(x,\tau)
R_{i\sigma}(x,\tau)\rangle.  
\label{SC}
\end{equation}

\subsection{Interchain Coupling}
\subsubsection{Coulomb Interaction}
As a source of the interchain coupling 
we first consider the Coulomb interaction
\cite{Saub,Nakajima-Okabe}.
The Coulomb interaction is 
introduced by adding the following action,  
\begin{eqnarray}
S_{\rm es} &=& -i \int{\rm d}\tau\, \int{\rm d}x\,
\sum_{i} 
\varphi_{i}(x,\tau) \rho_{i}(x,\tau)
\nonumber\\
&&+ \int {\rm d}\tau 
\int {\rm d}{\bf r}\,
\frac{1}{8 \pi}|\nabla \varphi({\bf r},\tau)|^{2},
\label{el-st-action}
\end{eqnarray}
where $\varphi_{i}(x,\tau)$ is the value of 
the scalar potential $\varphi({\bf r},\tau)$ 
on the $i$-th chain and $\rho_{i}(x,\tau)$ is 
the total charge density on the $i$-th chain. 
We write 
$\rho_{i}\,(x,\tau)=\rho_{i}^{\rm el}(x,\tau)
+\rho_{i}^{\rm ion}(x,\tau)$, 
where $\rho_{i}^{\rm el}(x,\tau)$ and 
$\rho_{i}^{\rm ion}(x,\tau)$ 
are the electronic and the ionic contribution 
to the charge density, respectively. 

The electronic contribution $\rho_{i}^{\rm el}$ 
has two characteristic parts, 
the long range ($q \sim 0$) and the 
short range ($q \sim 2k_{F}$) charge 
modulation, which we denote by $\rho_{0i}^{\rm el}$ and 
$\rho_{Qi}^{\rm el}$, respectively. 
These quantities are written as, 
\begin{eqnarray}
\rho _{i}^{\rm el} &=&\rho _{0i}^{\rm el}+\rho _{Qi}^{\rm el}, 
\nonumber \\
\rho _{0i}^{\rm el} &=& -e 
\left(R_{i\sigma }^{*}
R_{i\sigma} + L_{i\sigma}^{*} L_{i\sigma}
\right), 
\nonumber \\
\rho_{Qi}^{\rm el} &=& -e
\left(
L_{i\sigma}^{*} 
R_{i\sigma}
{\rm e}^{iQx}
+
R_{i\sigma}^{*} 
L_{i\sigma}
{\rm e}^{-iQx}
\right).
\label{rho-e}
\end{eqnarray}
The ionic contribution $\rho _{i}^{\rm ion}$ 
also has two
contributions $\rho_{0i}^{\rm ion}$ and 
$\rho_{Qi}^{\rm ion}$. 
For ions 
$\rho_{0i}^{\rm ion}$ is a constant,
since we take account of 
the phonons with wave numbers 
close to $\pm 2k_{F}$ only.
Noting that the
displacement of ions $u_{i}(x)$ is given 
from Eqs. (\ref{displacement}) and 
(\ref{delta}) by, 
\begin{equation}
u_{i}(x)=\frac{i\alpha _{Q}}{g}
\left\{ \Delta _{i}(x){\rm e}^{iQx}-
\Delta_{i}^{*}(x){\rm e}^{-i Q x}\right\}, 
\end{equation}
we obtain the ionic contribution 
to the charge density modulation as, 
\begin{eqnarray}
\rho_{0i}^{\rm ion} &=&\bar{\rho}, \\
\rho_{Qi}^{\rm ion} &=&-\bar{\rho}\,
\frac{\partial u_{i}}{\partial x},
\nonumber\\
&=&\frac{\bar{\rho}\alpha _{Q}Q}{g}
\left\{
\Delta _{i}(x){\rm e}^{i Q x}+
\Delta_{i}^{*}(x){\rm e}^{-iQx}
\right\},
\label{rho_Q}
\end{eqnarray}
where ${\bar{\rho}}$ is the 
average ionic charge density. 
Note that we neglected 
the derivative
$\partial_{x} \Delta _{i}(x)$ 
in Eq. (\ref{rho_Q}), 
since its 
effect is negligible as compared to 
$Q {\bar \Delta}_{i}$. 

As a whole, the total charge density $\rho_{0i}$ 
and $\rho_{Qi}$ are given by the 
summation of the electronic and the 
ionic contributions as, 
\begin{eqnarray}
\rho_{0i}(x)&=&\bar{\rho}
-e 
\left(
R_{i\sigma}^{*}
R_{i\sigma}+
L_{i\sigma}^{*}
L_{i\sigma}\right),
\\
\rho _{Qi} &=& A_{i}^{*}(x){\rm e}^{-iQx}+
A_{i}(x){\rm e}^{iQx}, 
\label{TotalCharge}
\end{eqnarray}
where
\begin{eqnarray}
A_{i}(x) &=&
\eta
\Delta_{i}(x)
-e 
L_{i\sigma}^{*}
R_{i\sigma}, 
\label{A}
\end{eqnarray}
with $\eta$ being $\bar{\rho}\alpha_{Q}Q/g$. 

In order to clarify the 
nature of the Coulomb interaction, 
first we trace out $\varphi(x,\tau)$. 
One can easily see that the following term 
is generated, 
\begin{eqnarray}
S_{\rm C}&=&\frac{1}{2}
\int_{0}^{\beta \hbar}{\rm d}\tau\, 
\sum_{ij} 
\int_{0}^{L}dx
\int_{0}^{L}dx^{\prime}\frac{\rho_{i}(x)
\rho_{j}(x^{\prime})}
{\sqrt{(x-x^{\prime})^{2}+d_{ij}^{2}}}, 
\nonumber\\
&&
\end{eqnarray}
where $d_{ij}$ is the distance between 
the $i$-th and the $j$-th chain and
the summation over $i$ and $j$ should 
be taken over all the chains. 
This expression can be rewritten using 
Fourier transformation as, 
\begin{eqnarray}
S_{{\rm C}} &=& 
\frac{1}{2}\int_{0}^{\beta \hbar}{\rm d}\tau\, 
\sum_{i,j}\frac{1}{L}\sum_{q}\tilde{v}_{ij}(q)
\tilde{\rho}_{i}(q)\tilde{\rho}_{j}(-q),
\label{coulomb1} \\
\tilde{\rho}_{i}(q) &=&\int_{0}^{L}dx\,\rho_{i}(x)\,
e^{-i q x},
\label{coulomb2} \\
\tilde{v}_{ij}(q) &=&\int_{0}^{L}dx\,
\frac{{\rm e}^{- i q x}}
{\sqrt{x^{2}+d_{ij}^{2}}},
\nonumber\\
&=& 2 K_{0}(d_{ij}|q|) \phantom{aaa}
(L \longrightarrow \infty), 
\label{coulomb3}
\end{eqnarray}
where $K_{0}(x)$ is the modified Bessel function. 
We assume that $d_{ii} \equiv d_{0} \neq 0$ in order 
to avoid divergence which occurs when $i=j$ in
Eq. (\ref{coulomb3}). 
The divergence arises from the fact that we considered 
the purely one-dimensional case, 
{\it i.e.}, infinitely thin chains. 
In reality, 
$d_{0}$ corresponds to the actual thickness of the chains, 
which may be given by the size of ions 
(or, more precisely, the extension of the electronic wave
function on an ion).

By substituting $\rho_{i}(x) = \rho_{0i}(x)+
\rho_{Qi}(x)$ 
into Eqs. (\ref{coulomb1}) and (\ref{coulomb2}), we obtain, 
\begin{eqnarray}
S_{\rm C}&=&
\frac{1}{2}
\int_{0}^{\beta \hbar}{\rm d}\tau\, 
\sum_{i,j}\frac{1}{L}
\sum_{q}\Bigl[\tilde{v}_{ij}(q)
\tilde{\rho}_{0i}(-q)\tilde{\rho}_{0j}(q)
\nonumber\\
&& + 2 \tilde{v}_{ij}(Q+q)\tilde{A}_{i}^{*}(q)
\tilde{A}_{j}(q)\Bigr],
\nonumber\\
&\equiv& S_{\rm C}^{(1)}+S_{\rm C}^{(2)},
\label{coulomb4}
\end{eqnarray}
where $\tilde{\rho}_{0i}(q)$ and 
$\tilde{A}_{i}(q)$ are the Fourier
components of $\rho _{0i}(x)$ and $A_{i}(x),$ respectively. 
We have neglected the cross term between 
$\rho_{0i}(x)$ and $\rho_{Qi}(x)$, since 
$R_{i \sigma}$ and $L_{i \sigma}$ 
are slowly varying. 

Note that the interactions in 
$S_{\rm C}^{(1)}$ and $S_{\rm C}^{(2)}$ 
have different range in $d_{ij}$. 
As is shown in Eq. (\ref{coulomb3}), 
$\tilde{v}_{ij}(q)$ is given by $2 K_{0}(d_{ij}|q|)$ 
which decays exponentially for $d_{ij}|q| 
\agt 1$. 
Since the wave number $q$ in $S_{\rm C}^{(1)}$ 
can take the value $q \sim 0$, 
$K_{0}(d_{ij}|q|)$ is not negligible even for 
large $d_{ij}$. 
In $S_{\rm C}^{(2)}$, on the other hand, 
$\tilde{v}_{ij}(Q+q)$ equals approximately to 
$2 K_{0}(Q d_{ij})$ and this 
is negligible for $d_{ij} \agt 1/Q = 1/(2 k_{F})$. 
The Fermi wavelength is usually the
order of the lattice constant and 
we can limit the summation with respect to $i$ and $j$ in 
$S_{\rm C}^{(2)}$ to adjacent chains only. 

We rewrite the long range part $S_{\rm C}^{(1)}$, 
restoring the scalar potential, as 
\begin{eqnarray}
S_{\rm C}^{(1)}&=&\int_{0}^{\beta \hbar }d\tau \,
\sum_{i}\int_{0}^{L}dx\,
\nonumber\\
&&\biggl[-i\varphi_{i}\left\{\bar\rho - 
e \left(
R_{i\sigma}^{*}
R_{i\sigma}+
L_{i\sigma}^{*}
L_{i\sigma}\right)
\right\}
\nonumber\\
&&+\frac{1}{8\pi}\biggl\{
(\partial _{x}\varphi _{i})^{2}+
\sum_{\alpha}
\frac{(\varphi _{i+\alpha}-
\varphi_{i})^{2}}{d_{\alpha}^{2}}\biggr\}\biggr],
\label{coulomb5}
\end{eqnarray}
where $\alpha = \{{\hat y},{\hat z}\}$ and 
$i+{\hat y}$ and $i+{\hat z}$ symbolically denote the 
chains shifted by one lattice vector 
in positive $y$- and $z$-direction 
from the $i$-th chain, respectively. 
Note that only the slowly varying part of 
the scalar potential with a wave vector 
much smaller than $k_{F}$ 
should be taken into account here. 

The short range part 
$S_{\rm C}^{(2)}$ can be divided into two 
kinds of terms: 
{\it intrachain} term ($j=i$)
and the {\it nearest-neighbor} term ($j=i+\alpha$).
The {\it intrachain} term 
has three parts: ion-ion interaction, 
electron-ion interaction and  
electron-electron interaction. 
The first and the second one are 
absorbed into the phonon energy $\hbar \omega_{Q}$ 
and the electron-phonon 
coupling constant $g$, 
respectively. 
In this paper we assume that 
the electron-electron interaction is 
small and negligible. 
Hence we suppress the intrachain term 
in the following. 

The {\it nearest-neighbor} term ($j=i+\alpha$) 
can be written, 
using the approximation,  
$\tilde{v}_{i i+\alpha}(Q+q)\rightarrow 
\tilde{v}_{i i+\alpha}(Q)
\equiv {\tilde v}_{\alpha}$, as, 
\begin{eqnarray}
S_{\rm C}^{(2)} &=& \int_{0}^{\beta \hbar}
{\rm d}\tau\, 
\sum_{i,\alpha }
{\tilde v}_{\alpha}
\int_{0}^{L}{\rm d} x\, 
\Bigl\{
A_{i}^{*}(x) A_{i+\alpha}(x) 
\nonumber\\
&&\phantom{aaaaa} + 
A_{i+\alpha}^{*}(x) A_{i}(x)
\Bigr\}.
\label{interchain1}
\end{eqnarray}

\subsubsection{Electron hopping}

Next we discuss the effect of 
electron hopping between chains. 
We consider only the hopping between 
adjacent chains and introduce the following term, 
\begin{eqnarray}
S_{\rm hop} &=& 
\int_{0}^{\beta\hbar} {\rm d} \tau\, 
\sum_{i,\alpha} 
\int_{0}^{L} {\rm d}x\, 
\nonumber\\
&&
\left\{t_{\alpha}
\left(
R_{i+\alpha \sigma}^{*} R_{i \sigma} +
L_{i+\alpha \sigma}^{*} L_{i \sigma}
\right)+{\rm c.c.}
\right\},
\label{hopping}
\end{eqnarray}
where $t_{\alpha}$ is transfer integral between 
the $i$-th and the $(i+\alpha)$-th chains 
and c.c. stands for the complex conjugate. 

\subsection{Chiral Transformation}

In integrating out the electronic degree of 
freedom, it is convenient to employ 
chiral transformation so that 
the phase $\theta_{i}(x)$ is eliminated from 
$\Delta_{i}(x)$. 
In this paper we study the 
temperature region not too close to $T_{c}$. 
Hence the amplitude of the order 
parameter can be treated as a constant. 
We write 
$\Delta _{i}(x) = 
{\bar{\Delta }}_{i}{\rm \ e}^{i\theta _{i}(x)}$ 
in the following. 
The chiral transformation is introduced as 
follows: 
\begin{equation}
R_{i\sigma} \longrightarrow 
\exp(i \theta_{i}/2) R_{i\sigma},
\phantom{a }
L_{i\sigma} \longrightarrow 
\exp(-i \theta_{i}/2) L_{i\sigma}. 
\label{chiral}
\end{equation}
By this transformation $S_{\mbox{\scriptsize e-p}}$ is 
changed to, 
\begin{eqnarray}
S_{\mbox{\scriptsize e-p}}&=&\int_{0}^{\hbar \beta } d\tau
\sum_{i}\int_{0}^{L}dx\left[{\Psi}_{i\sigma}^{\dagger}\cdot 
{\bf K}_{i}\cdot {\Psi}_{i\sigma }+
\frac{\hbar \omega _{Q}}{2g^{2}}\bar{\Delta}_{i}^{2}\right], 
\nonumber\\&&
\label{transformed-action}
\end{eqnarray}
where 
\begin{eqnarray}
{\bf K}_{i}(x,\tau ) &=&\left( 
\begin{array}{cc}
\hat{k}_{0i}^{(+)}+\hat{k}_{1i} & {\bar{\Delta }}_{i} \\ 
{\bar{\Delta }}_{i} & \hat{k}_{0i}^{(-)}+\hat{k}_{1i}
\end{array}
\right),  \label{action-3} \\
\hat{k}_{0i}^{(\pm )} &=& \hbar \partial_{\tau} \mp i\hbar
v_{F} \partial_{x},  \label{action-4} \\
\hat{k}_{1i} &=&\frac{\hbar ^{2}}{2m}\left\{k_{F}
\partial_{x} \theta _{i}
+\frac{1}{4}\left(\partial_{x} \theta _{i}
\right) ^{2}\right\},
\label{K}
\label{k1i}
\end{eqnarray}
and $v_{F} = \hbar k_{F}/m$ is the Fermi velocity. 

In Eq. (\ref{action-4}) we have neglected 
the term proportional to $\partial^{2}_{x}$. 
As pointed out by Ishikawa and Takayama 
\cite{Ishikawa-Takayama}  
this term is necessary to obtain 
the free energy which describes 
electromagnetic response of the system correctly. 
On the other hand, if one neglects this term and 
approximates the dispersion by a linear one,  
so-called {\it chiral anomaly} term 
must be taken into account
\cite{Su-Sakita,Sakita-Shizuya,Nagaosa-Oshikawa}. 
In this paper, 
we take the latter way and add the following term to 
our action, 
\begin{equation}
S_{\rm ch}= \int_{0}^{\beta \hbar}
{\rm d}\tau\,
\sum_{i}\int_{0}^{L}dx\ 
\frac{i e}{\pi }\partial _{x}\theta _{i}(x)\varphi_{i}(x).  
\label{CA}
\end{equation}

The terms $S_{\rm C}^{(2)}$ and 
$S_{\rm hop}$ are transformed as, 
\begin{eqnarray}
S_{\rm C}^{(2)}&\longrightarrow& \int_{0}^{\beta \hbar}
{\rm d}\tau\, 
\sum_{i,\alpha }
{\tilde v}_{\alpha}
\int_{0}^{L}{\rm d} x\, 
\nonumber\\
&&\biggl[
2 \eta^{2}{\bar\Delta}_{i}
{\bar\Delta}_{i+\alpha} \cos (\theta_{i+\alpha} - \theta_{i})
\nonumber\\
&&\phantom{\biggl[}-\eta e 
\Bigl\{{\bar\Delta}_{i}
\left(
R_{i+\alpha \sigma}^{*}
L_{i+\alpha \sigma} 
{\rm e}^{-i(\theta_{i+\alpha} - \theta_{i})}+
{\rm c.c.}
\right)
\nonumber\\
&&\phantom{aaaaaaaaaaa} + (i \longleftrightarrow i+\alpha)
\Bigr\}
\nonumber\\
&&\phantom{\biggl[}
+e^{2}\Bigl\{
R_{i \sigma}^{*}
L_{i \sigma} 
L_{i+\alpha \sigma'}^{*}
R_{i+\alpha \sigma'} 
{\rm e}^{i(\theta_{i+\alpha} - \theta_{i})}
\nonumber\\
&&\phantom{aaaaaaa}
+{\rm c.c.}
\Bigr\}
\biggr],
\label{Sc2}
\\
S_{\rm hop} 
&\longrightarrow&
\int_{0}^{\beta\hbar} {\rm d} \tau\, 
\sum_{i,\alpha} 
\int_{0}^{L} {\rm d}x\, \Bigl\{t_{\alpha}
\Bigl(
{\rm e}^{-i\frac{\theta_{i+\alpha}-\theta_{i}}{2}}
R_{i+\alpha \sigma}^{*} R_{i \sigma}
\nonumber\\
&&
+{\rm e}^{i\frac{\theta_{i+\alpha}-\theta_{i}}{2}}
L_{i+\alpha \sigma}^{*} L_{i \sigma}
\Bigr)+{\rm c.c}.
\Bigr\}.
\label{Shop}
\end{eqnarray}

\section{Derivation of the GL free energy}

The effective free energy of the condensate can be
obtained by integrating out the electronic 
degree of freedom in the following way,  
\begin{equation}
F[\theta _{i},\varphi _{i},{\bar\Delta}_{i}]=
-\frac{1}{\beta}
\ln \int \prod_{i}
{\mathcal D}[{\Psi}_{i\sigma }^{\dagger},
{\Psi}_{i\sigma }]\,
e^{- S_{\rm tot}/\hbar},  
\label{free-energy1}
\end{equation}
where $S_{\rm tot}$ is the total action. 
In performing this path integral, 
the following perturbative treatment  
is available,
\begin{eqnarray}
Z(\beta)&=&\int {\cal D}
[\Psi_{i\sigma}^{\dagger},\Psi_{i\sigma}]
\exp 
\left[-\{S_{0}+\delta S\}/\hbar\right],
\nonumber\\
&=& \int {\cal D}
[\Psi_{i\sigma}^{\dagger},\Psi_{i\sigma}]
\sum_{n=0}^{\infty}
\frac{(-1)^{n}\delta S^{n}}{n! \hbar^{n}}
{\rm e}^{-S_{0}/\hbar},
\nonumber\\
&=& Z_{0}(\beta)\times\sum_{n=0}^{\infty}
\frac{(-1)^{n} 
\left\langle \delta S^{n}\right\rangle_{c}}
{n! \hbar^{n}},
\nonumber\\
&=& Z_{0}(\beta)\times\exp\Bigl[-
\sum_{n=0}^{\infty}
\left\langle \delta S^{n}\right\rangle_{c}
\Bigr], 
\nonumber\\
&=& Z_{0}(\beta)\times
\exp\left[W\{\theta_{i},\varphi_{i},
{\bar \Delta}_{i}\} \right], 
\end{eqnarray}
where $S_{0}$ and $\delta S$ 
are the unperturbed and perturbed part, 
respectively. 
The bracket 
$\left\langle \cdots \right\rangle_{c}$ 
indicates that only the 
\lq\lq connected diagrams\rq\rq\ should 
be taken into account and 
$Z_{0}(\beta)$ is the 
unperturbed partition function. 
From this formula, 
the effective free energy $F$ is given by
$F = F_{0} - W/\beta$ 
where $F_{0} = - \ln(Z_{0}(\beta))/\beta$. 
From now on we assume that the amplitude 
of the order parameter $\bar{\Delta}_{i}$ 
does not depend on the chain index 
and write $\bar{\Delta}_{i}=\bar{\Delta}$. 

Based on the above mentioned method, 
the effective free energy can be derived. 
Here, for simplicity, 
we discuss only the resulting 
free energy, leaving 
the details of the calculations 
for Appendix \ref{int}. 
The free energy consists of 
two parts, the intrachain part 
$F_{\rm intra}$ and the
interchain part $F_{\rm inter}$. 
The intrachain part is given by,
\begin{eqnarray}
F_{\rm intra} &=& 
\sum_{i}\int_{0}^{L}dx
\biggl[
\frac{\hbar v_{F}}{4\pi }f_{{\rm s}}
\left(\partial_{x}\theta_{i}\right)^{2}
\nonumber\\
&& \phantom{aaaaa}+
\frac{e^{2}(1-f_{{\rm s}})}{\pi \hbar v_{F}}
\varphi_{i}^{2}+
\frac{i e f_{{\rm s}}}{\pi}\partial_{x}
\theta_{i}\varphi_{i}\biggr],
\label{elastic}
\end{eqnarray}
where the condensate fraction 
$f_{{\rm s}}$ behaves as, 
\begin{eqnarray}
f_{{\rm s}}
&=& \left\{
\begin{array}{ll}
\frac{1}{4\pi^{2}}
\beta^{2}\bar{\Delta}^{2}\zeta(3,\frac{1}{2}) & 
\mbox{\,\,\,\,for $\beta \bar{\Delta} \ll 0$}\\
1-\sqrt{2\pi \beta \bar{
\Delta }}\exp({-\beta \bar{\Delta}}) &
\mbox{\,\,\,\,for $\beta \bar{\Delta} \gg 0$}
\end{array}
\right.
\end{eqnarray}
with  
$\zeta(3,1/2)$ being the zeta function. 
(See Appendix \ref{gl} for details.)
The interchain part $F_{\rm inter}$ 
is given by,
\begin{equation}
F_{\rm inter} = \mbox{const.} + 
\sum_{i,\alpha}\int {\rm d}x\, 
\hbar \Gamma_{\alpha}\cos (\theta_{i+\alpha}- \theta_{i}), 
\label{inter-chain}
\end{equation}
where 
$\Gamma_{\alpha}$ is the interchain coupling constant 
given by, 
\begin{eqnarray}
\Gamma_{\alpha} &=& 
\frac{2\, \tilde{v}_{\alpha}
\bar{\Delta}^{2}}{\hbar}
\left(\frac{\bar{\rho}\alpha_{Q}Q}{g}
+ \frac{e \hbar \omega_{Q}}{2 g^{2}}
\right)^{2} + 
\frac{|t_{\alpha}|^{2}}{\pi \hbar^{2} v_{F}}
f_{{\rm s}}.
\nonumber\\
&&
\label{gamma1}
\end{eqnarray}
As one can see from Eqs. (\ref{inter-chain}) and 
(\ref{gamma1}), the CDW prefers 
the phase difference of the neighboring chains 
to be $\pi$ since $\Gamma_{\alpha}$ is positive. 
In this paper we assume that the chains form a 
bipartite crystal and shift the phase of the
chains on one sublattice by $\pi$. 
In this new notation $\Gamma_{\alpha}$ is changed to 
$-\Gamma_{\alpha}$, leading to the 
following expression for 
$F_{\rm inter}$, 
\begin{eqnarray}
F_{\rm inter}   
&=& \mbox{const.}
-\sum_{i,\alpha} \int {\rm d}x\, 
\hbar \Gamma_{\alpha}\cos 
(\theta_{i+\alpha} - \theta_{i}),
\nonumber\\
&\approx& F_{\rm inter}^{(0)} + 
\sum_{i,\alpha} \int {\rm d}x\, 
\frac{\hbar \Gamma_{\alpha}}{2}
(\theta_{i+\alpha} - \theta_{i})^{2}. 
\end{eqnarray}
where $F_{\rm inter}^{(0)}$ is 
the change of the condensation energy 
due to interchain coupling. 

Next we take the continuum limit 
with respect to the chain index 
$i$ by writing $\theta({\bf r})$ 
instead of $\theta_{i}(x) $. 
The total free energy becomes,
\begin{eqnarray}
F &=& F^{(0)}+\int {\rm d}{\bf r}
\biggl[\frac{K}{2}
\left\{(\partial_{x}\theta)^{2}+
\gamma_{y}^{2}(\partial_{y}\theta)^{2}+
\gamma_{z}^{2}(\partial_{z}\theta)^{2}\right\}
\nonumber\\
&&+ i\, e\, J\, \varphi\, \partial _{x}\theta +
\frac{1}{8\pi}\left\{|\nabla \varphi|^{2}+
(\lambda_{0}^{-2}+\lambda_{\rm res}^{-2})\varphi^{2}
\right\} 
\nonumber\\
&&+ i\,e\,\varphi\, \rho^{\rm ext}\biggr],
\label{free-energy}
\end{eqnarray}
where $K= \hbar v_{F} N_{\bot} f_{\rm s}/(2 \pi)$ and 
$J = N_{\bot} f_{\rm s}/\pi$ 
with $N_{\bot}$ being the cross sectional density of the chains. 
The anisotropy parameters $\gamma_{y}$ and $\gamma_{z}$ are 
given by 
$K \gamma_{\alpha}^{2} = \hbar \Gamma_{\alpha} d_{\alpha}^{2}$ 
($\alpha = y,z$). 
Here we have included the energy of electric field, 
{\it i.e.}, the last two terms of Eq. (\ref{coulomb5}) and 
also introduced the external charge 
$\rho^{\rm ext}({\bf r})$ which is a source of the external 
electric field. 

The first and the second term of 
$\Gamma_{\alpha}$ in Eq. (\ref{gamma1}) 
are proportional to 
${\bar \Delta}^{2}$ and $f_{\rm s}$, 
respectively. 
As is known for the case of superconductivity, 
these quantities, namely the square of the gap 
and the superfluid density, have different temperature 
dependence at $T \ll T_{c}$, 
although the relation $f_{\rm s} \propto 
{\bar \Delta}^{2}$ holds near $T_{c}$ 
\cite{Tinkham}. 
Therefore $\gamma_{\alpha}^{2}$, 
which is proportional to 
$\Gamma_{\alpha}/f_{\rm s}$, 
has a weak temperature dependence 
at $T \ll T_{c}$ 
arising from the difference between 
${\bar \Delta}^{2}$ and $f_{\rm s}$. 
This temperature dependence, however, 
is almost negligible because the temperature 
dependence of ${\bar \Delta}^{2}$ and $f_{\rm s}$ 
are both exponentially suppressed 
at low temperatures. 

In Eq. (\ref{free-energy}) 
$\lambda_{0}$ is the electrostatic screening 
length by the quasiparticles, 
which is given by 
$\lambda_{0}^{-2} = \lambda_{\rm TF}^{-2}(1 - f_{\rm s})$ 
where $\lambda_{\rm TF}^{-2} = 
8 e^{2} N_{\bot} /(\hbar v_{F})$
is the Thomas-Fermi screening length of the individual 
1-D chain. 
At $T=0$, $f_{\rm s}$ tends to $1$ and 
$\lambda_{0}$ diverges. 
This is because we have assumed the complete nesting 
in our calculation. 
In actual CDW's, there exists residual carriers 
which does not contribute to CDW ordering even at $0K$. 
Such materials are called \lq\lq metallic\rq\rq\, CDW's. 
In this case, the divergence of $\lambda_{0}$ is suppressed. 
We have introduced 
$\lambda_{\rm res}$ to incorporate 
the contribution of these carriers. 

The amplitude fluctuation of the order parameter can be 
treated in a similar way as we have treated the 
phase fluctuation in this paper. 
For purely one-dimensional systems 
this was carried out by 
Brazovski\u i and Dzyaloshinski\u i
\cite{Brazovskii-Dzyaloshinskii}. 
The extension of the present calculation to the case with 
the amplitude mode may be straightforward. 

\section{Summary}
In this paper we have derived the effective 
free energy of the phase of CDW employing the 
path integral method. 
Especially, a careful treatment 
of the interchain coupling, {\it i.e.}, 
Coulomb interaction and electron hopping, was presented. 
We have clarified the followings: 
\begin{enumerate}
\item The Coulomb interaction is divided into two 
parts, which correspond to 
slow modulation ($q \sim 0$) 
and fast oscillation 
($q \sim 2 k_{F}$) of charge density. 
The former determines the large scale 
electromagnetic response of the CDW's and 
the latter plays an important role in the 
elastic coupling between CDW's on adjacent 
chains. 
\item The electron hopping between neighboring 
chains also contributes to the 
elastic interchain coupling of CDW. 
\item The contribution of the Coulomb interaction 
and the electron hopping to the interchain coupling 
have slightly different temperature dependence 
at low temperatures. 
As a result the anisotropy parameters also become 
weakly temperature dependent at low temperatures, 
although its dependence is negligibly small. 
\item The quasiparticles give a finite electrostatic 
screening length. 
The length is given by the Thomas-Fermi length 
above $T_{c}$ but is enhanced in lower temperatures 
due to the suppression of the density of states 
by the energy gap of CDW. 
In \lq\lq semiconducting\rq\rq\,
CDW's, the quasiparticle screening length  
is divergent at $T=0$ but it remains finite 
in \lq\lq metallic\rq\rq\, ones. 
\end{enumerate}

\acknowledgments

This work was supported by 
a Grant-In-Aid for Scientific 
Research (10740159 and 11740196) from the Ministry of Education, 
Science and Culture, Japan. 
The authors (M.H. and H.Y.) are grateful to 
\lq\lq Stichting voor Fundamenteel Onderzoek der Materie\rq\rq 
(FOM) for supporting during their stay in 
Delft University of Technology. 
They also thank Prof. G. E. W. Bauer for 
kind hospitality and useful discussions. 

\appendix

\section{Integration over electronic field}
\label{int}

From Eqs. (\ref{coulomb5}), (\ref{transformed-action}), 
(\ref{CA}), 
(\ref{Sc2}) and (\ref{Shop}), we obtain the total 
action of the system. 
For simplicity we set ${\bar \Delta_{i}}\rightarrow 
{\bar \Delta}$ in the following. 
We divide the total action into two parts, 
unperturbed part $S_{0}$ and 
perturbed part $\delta S$. 
Here we assign 
$S_{0}$ the part of $S_{\mbox{e-p}}$ in 
Eq. (\ref{transformed-action}) which does not 
include $\theta_{i}$. 
The rest of the terms are assigned to $\delta S$. 

The unperturbed action is given as follows, 
\begin{eqnarray}
\frac{S_{0}}{\hbar} &=& 
\int_{0}^{\beta\hbar}{\rm d}\tau\,{\rm d}\tau'\,
\sum_{i}
\int_{0}^{L} {\rm d}x\, {\rm d}x'\,
\nonumber\\
&&\Psi_{i\sigma}^{\dagger}(x,\tau)
\cdot \{-
{\cal G}(x-x',\tau-\tau')\}^{-1}
\cdot \Psi_{i\sigma}(x',\tau'),
\nonumber\\
&=& \sum_{i}
\sum_{k,\varepsilon_{n}}
\tilde\Psi_{i\sigma}^{\dagger}(k,i \varepsilon_{n})\cdot
\{-\tilde{\cal G}(k,i \varepsilon_{n})\}^{-1}
\cdot \tilde\Psi_{i\sigma}(k,i \varepsilon_{n}),
\nonumber\\
\end{eqnarray}
where,
\begin{eqnarray}
\tilde{\cal G}(k,i \varepsilon_{n}) 
&=&
\left(
\begin{array}{cc}
\tilde{\cal G}^{R}(k,i \varepsilon_{n}) & 
\tilde{\cal F}(k,i \varepsilon_{n})\\
\tilde{\cal F}(k,i \varepsilon_{n}) & 
\tilde{\cal G}^{L}(k,i \varepsilon_{n})
\end{array}
\right),
\end{eqnarray}
and the Green functions are given by 
\begin{eqnarray}
\tilde{\cal G}^{R}(k,i \varepsilon_{n}) &=&
\frac{-\hbar \left(
i \hbar \varepsilon_{n} + \xi\right)}
{(\hbar \varepsilon_{n})^{2} + E_{i}(k)^{2}},
\\
\tilde{\cal G}^{L}(k,i \varepsilon_{n}) &=&
\frac{-\hbar \left(
i \hbar \varepsilon_{n} - \xi\right)}
{(\hbar \varepsilon_{n})^{2} + E_{i}(k)^{2}},
\\
\tilde{\cal F}(k,i \varepsilon_{n}) &=&
\frac{-\hbar {\bar \Delta}}
{(\hbar \varepsilon_{n})^{2} + E(k)^{2}},
\label{green}
\end{eqnarray}
with  
$E(k) = \sqrt{\xi^{2} + \bar{\Delta}^{2}}$, 
$\xi = \hbar v_{F} k$ and $\varepsilon_{n} = (2 n -1)\pi/
(\beta\hbar)$. 

The perturbed term $\delta S$ 
is given as follows, 
\begin{eqnarray}
\delta S &=& S_{1} + S_{2} +
S_{\rm C}^{(2)} + S_{\rm hop},
\nonumber\\
S_{1} &=& \int
{\rm d} \tau\, 
\sum_{i} 
\int {\rm d} x\,
\chi_{i}(x) \left(
R_{i \sigma}^{*} R_{i \sigma}
+ L_{i \sigma}^{*} L_{i \sigma}
\right), 
\\
S_{2} &=& -i \int d\tau \,
\sum_{i}\int dx\,
\varphi_{i}\left\{\bar\rho
-\frac{e}{\pi}\partial_{x}\theta_{i}
\right\},
\label{a7}
\end{eqnarray}
where $\chi_{i}(x)$ is defined by 
\begin{eqnarray}
\chi_{i} (x) &\equiv&
\frac{i\,e}{\hbar}\varphi_{i} +
\frac{1}{2}v_{F}\partial_{x}\,
\theta _{i}
+\frac{\hbar}{8m}(\partial _{x}\theta_{i})^{2}. 
\label{chi}
\end{eqnarray}
$S_{\rm C}^{(2)}$ and $S_{\rm hop}$ are 
given by Eq. (\ref{Sc2}) and 
Eq. (\ref{Shop}), respectively. 
The energy of the electric field, 
{\it i.e.}, 
the last two terms of Eq. (\ref{coulomb5}) 
is not included here. 

After the integration with respect to the electronic 
degree of freedom we obtain the following results: 
\begin{eqnarray}
\left\langle S_{1} \right\rangle_{c}
&=& \int {\rm d}\tau\,
\sum_{i}
\int {\rm d} x\, 
n_{\rm e}\,\chi_{i}(x),
\label{S1}
\end{eqnarray}
where $n_{\rm e}$ is the equilibrium electron density 
on a chain which is assumed to be a constant 
independent of $i$ 
and given by $2 k_{F}/\pi$. 
In Eq. (\ref{S1}), 
the first term of $\chi_{i}(x)$ 
should be cancelled 
by the ionic background charge, 
{\it i.e.}, the first term of Eq. (\ref{a7}) 
for charge neutrality. 
In the present calculations, the cancellation of 
second term of $\chi_{i}(x)$, which is 
proportional to $\partial_{x}\theta_{i}$, 
is not complete. 
This term should vanish in order to maintain 
the stability of CDW with a wave number $2 k_{F}$. 
We consider that this cancellation is 
fulfilled if we take into account the 
$\partial_{x}^{2}$-term 
which we have neglected in 
Eq. (\ref{action-3}).  

The second order term with respect to 
$S_{1}$ is given as follows, 
\begin{eqnarray}
\left\langle \{S_{1}\}^{2} \right\rangle_{c}
= &&
\sum_{i}
\int {\rm d} X\, {\rm d} X'\, 
\chi_{i}(X) \chi_{i}(X') 
\nonumber\\
&&\times
2 \Bigl\{-
{\cal G}^{R}(X-X')
{\cal G}^{R}(X'-X)
\nonumber\\
& &\phantom{2 \Bigl\{}
- {\cal G}^{L}(X-X')
{\cal G}^{L}(X'-X) 
\nonumber\\
& &\phantom{2 \Bigl\{}
- 2 {\cal F}(X-X')
{\cal F}(X'-X)
\Bigr\},
\end{eqnarray}
where we have introduced $X \equiv (x,\tau)$
for simplicity. 
Since $\chi_{i}(x)$ is slowly varying 
as compared to 
${\cal G}^{R}$, ${\cal G}^{L}$ 
or ${\cal F}$, 
we can use the following treatment, 
\begin{eqnarray}
&&{\cal G}^{R}(X-X')
{\cal G}^{R}(X'-X)
\nonumber\\
&&\phantom{aaaaa}
\longrightarrow
\delta(X-X')\times
\int {\rm d}X\, 
{\cal G}^{R}(X)
{\cal G}^{R}(-X)
\nonumber\\
&&\phantom{aaaaa}
= \delta(X-X')\times
\frac{1}{\beta\hbar L}
\sum_{k,\varepsilon_{n}}
\tilde{\cal G}^{R}(k,i \varepsilon_{n})^{2}.
\end{eqnarray}
This treatment corresponds to 
GL expansion. 
As we see in the next section, 
the following quantities are 
rewritten in terms of the 
the condensate fraction $f_{\rm s}$ as, 
\begin{eqnarray}
&&\frac{1}{\beta\hbar L}
\sum_{k,\varepsilon_{n}}
\left\{
\tilde{\cal G}^{R}(k,i \varepsilon_{n})^{2}+
\tilde{\cal G}^{L}(k,i \varepsilon_{n})^{2}
\right\}
\nonumber\\
&&\phantom{aaa}
=-\frac{1}{\pi v_{F}}
\left(1 - \frac{f_{\rm s}}{2}\right),
\label{gg}\\
&&\frac{1}{\beta\hbar L}
\sum_{k,\varepsilon_{n}}
\tilde{\cal F}(k,i \varepsilon_{n})^{2}
=\frac{1}{4 \pi v_{F}} f_{\rm s}, 
\label{ff}
\end{eqnarray}
which yields the following expression, 
\begin{eqnarray}
&-&\frac{1}{\hbar}
\left\langle S_{1} \right\rangle_{c}
+ \frac{1}{2 \hbar^{2}}
\left\langle \{S_{1}\}^{2} \right\rangle_{c}
-\frac{1}{\hbar}S_{2}
\nonumber\\
&=&
\int {\rm d} \tau\, 
\sum_{i}
\int {\rm d} x\, 
\biggl\{
- f_{\rm s}\frac{v_{F}}{4 \pi} 
(\partial_{x} \theta_{i})^{2} 
\nonumber\\
&&
- f_{\rm s}\frac{i e}{\pi\hbar}
\varphi_{i} \partial_{x} \theta_{i}
-(1 - f_{\rm s})\frac{e^{2}}{\pi v_{F} \hbar^{2}}
\varphi_{i}^{2}
\biggr\},
\nonumber\\
&\equiv& W_{1}.
\label{intraW}
\end{eqnarray}

The terms which includes two chains are given as follows, 
\begin{eqnarray}
\left\langle S_{\rm C}^{(2)}\right\rangle_{c} &=&
\sum_{i,\alpha} {\tilde v}_{\alpha}
\int_{0}^{\beta \hbar}{\rm d}\tau\,
\int {\rm d}x\, 
\left(
\frac{{\bar \rho} \alpha_{Q}Q}{g}
+ \frac{e \hbar \omega _{Q}}{2 g^{2}}
\right)^{2} \nonumber\\
&&
\times
2 {\bar \Delta}^{2} \cos(\theta_{i+\alpha}-
\theta_{i}), 
\end{eqnarray}
where we have used the relation, Eq. (\ref{SC}). 
We also obtain for the hopping term, 
\begin{eqnarray}
\left\langle \{S_{\rm hop}\}^{2}\right\rangle_{c} &=&
4 \int {\rm d}X\, {\rm d}X'\, 
\sum_{i, \alpha} |t_{\alpha}|^{2}
\Bigl\{
\nonumber\\
&&-{\cal G}^{R}(X-X'){\cal G}^{R}(X'-X)
\nonumber\\
&&-{\cal G}^{L}(X-X'){\cal G}^{L}(X'-X)
\nonumber\\
&&-{\cal F}^{*}(X-X'){\cal F}(X'-X)
{\rm e}^{-i (\theta_{i + \alpha} - \theta_{i})}
\nonumber\\
&&- {\cal F}_{i}(X-X'){\cal F}^{*}_{i}(X'-X)
{\rm e}^{i (\theta_{i + \alpha} - \theta_{i})}
\Bigr\},
\nonumber\\
&=& 4 \int {\rm d}X\, 
\sum_{i,\alpha}
|t_{\alpha}|^{2} \Bigl\{
\frac{1}{\pi v_{F}}
\left(
1 - \frac{f_{\rm s}}{2}
\right)
\nonumber\\
&&- \frac{f_{\rm s}}{2 \pi v_{F}}
\cos(\theta_{i + \alpha} - \theta_{i})
\Bigr\}.
\end{eqnarray}

Collecting these terms 
we obtain the interchain terms as, 
\begin{eqnarray}
&-&\frac{1}{\hbar}
\left\langle S_{\rm C}^{(2)}\right\rangle_{c} 
+ \frac{1}{2 \hbar^{2}}
\left\langle \{S_{\rm hop}\}^{2}\right\rangle_{c}
\nonumber\\
&=& \int {\rm d}\tau\, 
\sum_{i, \alpha}
\int {\rm d}x\,
\biggl\{\frac{2 |t_{\alpha}|^{2}}{\hbar^{2} \pi v_{F}}
\left(1 - \frac{f_{\rm s}}{2}\right)
\nonumber\\
&&\phantom{aaaaaazaaaaa}
- \Gamma_{\alpha}
\cos (\theta_{i+\alpha} - \theta_{i})\biggr\},
\nonumber\\
&\equiv& W_{2}.
\end{eqnarray}
where 
\begin{eqnarray}
\Gamma_{\alpha} &=& 
\frac{2\, \tilde{v}_{\alpha}
\bar{\Delta}^{2}}{\hbar}
\left(\frac{\bar{\rho}\alpha_{Q}Q}{g}
+ \frac{e \hbar \omega_{Q}}{2 g^{2}}
\right)^{2} 
+ \frac{|t_{\alpha}|^{2}}{\pi \hbar^{2} v_{F}}
f_{\rm s}.
\label{gamma}
\label{interW}
\end{eqnarray}
Here we assumed $\theta_{i}$ and $\varphi_{i}$
to be classical variables and 
neglected their $\tau$-dependence. 

The intrachain part $F_{\rm intra}$ and 
the interchain part $F_{\rm inter}$ of 
the free energy 
are given, respectively, by 
$F_{\rm intra} = - W_{1}/\beta$ and 
$F_{\rm inter} = - W_{2}/\beta$. 

\section{Calculation of the GL coefficients}
\label{gl}

In this section we derive 
Eq. (\ref{gg}) and (\ref{ff}). 
We introduce a cut off energy, 
$\hbar v_{F} |k| \sim \Lambda$, 
which should be set to 
$\infty$ in the end of the calculation. 

Eq. (\ref{gg}) can be obtained in the following way,
\begin{eqnarray}
&&\frac{1}{\beta \hbar L}
\sum_{k,\omega_{n}}
\Bigl\{
{\cal G}^{R}(k,i\varepsilon_{n})^{2}+ 
{\cal G}^{L}(k,i\varepsilon_{n})^{2}\Bigr\}
\nonumber\\
&&=\frac{2}{\beta \hbar}
\sum_{\varepsilon_{n}}
\frac{\hbar}{\pi v_{F}}
\int_{0}^{\Lambda}{\rm d}\xi\,
\frac{-(\hbar \varepsilon_{n})^{2}+\xi^{2}}
{\{(\hbar \varepsilon_{n})^{2}+E(k)^{2}
\}^{2}},
\nonumber\\
&&=\frac{2}{\beta \hbar}
\sum_{\varepsilon_{n}}
\frac{\hbar}{\pi v_{F}}
\int_{0}^{\Lambda}{\rm d}\xi\,
\biggl[
\frac{\xi^{2}-(\hbar \varepsilon_{n})^{2}-
\bar{\Delta}^{2}}
{\{(\hbar \varepsilon_{n})^{2}+E(k)^{2}
\}^{2}}
\nonumber\\
&&\phantom{aaaaaaaaaaaaaaaa}
+ \frac{
\bar{\Delta}^{2}}
{\{(\hbar \varepsilon_{n})^{2}+E(k)^{2}
\}^{2}}
\biggr],
\nonumber\\
&&\equiv I_{1}+I_{2}.
\label{glcgg}
\end{eqnarray}
$I_{1}$ is calculated as, 
\begin{eqnarray}
I_{1} &=& \frac{1}{\beta \hbar}
\sum_{\varepsilon_{n}}\frac{\hbar}{\pi v_{F}}
\left(
\frac{-\Lambda}{\Lambda^{2}+(\hbar \varepsilon_{n})^{2} +
\bar{\Delta}^{2}}
\right),
\nonumber\\
&=& 
\frac{-1}{\pi v_{F}}\frac{\Lambda}{\sqrt{\Lambda^{2}+
\bar{\Delta}^{2}}}\tanh
\frac{\beta \sqrt{\Lambda^{2}+
\bar{\Delta}^{2}}}{2},
\nonumber\\
&\longrightarrow& -\frac{1}{\pi v_{F}}  
\phantom{aaaa}(\Lambda \longrightarrow \infty),
\label{i1}
\end{eqnarray}
using the identity, 
\begin{equation}
\tanh z = z \sum_{n=-\infty}^{\infty}
\frac{1}{\{(2 n -1) \pi/2\}^{2}+z^{2}}.
\end{equation}
$I_{2}$ is also calculated in a similar way as, 
\begin{eqnarray}
I_{2} &=& \frac{\bar{\Delta}^{2}}{\beta \hbar}
\sum_{\varepsilon_{n}}\frac{2 \hbar}{\pi v_{F}}
\nonumber\\
&&\times\biggl[
\frac{\Lambda}{2 \{(\hbar \varepsilon_{n})^{2}+
\bar{\Delta}^{2}\}
\{(\hbar \varepsilon_{n})^{2}+
\bar{\Delta}^{2}+ \Lambda^{2}
\}} 
\nonumber\\
&&\phantom{a}
+ \frac{1}{2 
\sqrt{(\hbar \varepsilon_{n})^{2}+
\bar{\Delta}^{2}}^{3}}
\arctan \frac{\Lambda}
{\sqrt{(\hbar \varepsilon_{n})^{2}+
\bar{\Delta}^{2}}}
\biggr],
\nonumber\\
&=&
\frac{\bar{\Delta}^{2}}{2 \pi v_{F}}
\frac{1}{\Lambda}
\biggl(
\frac{1}{\bar{\Delta}}
\tanh \frac{\beta\bar{\Delta}}{2}
\nonumber\\
&&\phantom{aaaaaaa}
- \frac{1}{\sqrt{\Lambda^{2}+
\bar{\Delta}^{2}}}
\tanh \frac{\beta\sqrt{\Lambda^{2}+
\bar{\Delta}^{2}}}{2}
\biggr)
\nonumber\\
&&+
\frac{\bar{\Delta}^{2}}{\pi v_{F} \beta}
\sum_{\varepsilon_{n}}
\{(\hbar \varepsilon_{n})^{2}+
\bar{\Delta}^{2}\}^{-\frac{3}{2}} 
\nonumber\\
&&\phantom{aaaaaa}
\times\arctan
\frac{\Lambda}
{\sqrt{(\hbar \varepsilon_{n})^{2}+
\bar{\Delta}^{2}}},
\nonumber\\
&\longrightarrow & 
\frac{1}{\pi v_{F}}
\frac{\bar{\Delta}^{2}}{\beta}
\sum_{\varepsilon_{n}}
\{(\hbar \varepsilon_{n})^{2}+
\bar{\Delta}^{2}\}^{-\frac{3}{2}} 
\nonumber\\
&&\phantom{a}
\times\arctan
\frac{\Lambda}
{\sqrt{(\hbar \varepsilon_{n})^{2}+
\bar{\Delta}^{2}}}
\phantom{aa}(\Lambda \longrightarrow \infty).
\label{i2}
\end{eqnarray}

We define the condensate fraction 
$f_{\rm s}$ by 
\begin{eqnarray}
f_{\rm s} &=& 2 \pi v_{F} I_{2}
\nonumber\\
&=& \frac{4}{\beta}
\sum_{\varepsilon_{n}}\int_{0}^{\Lambda} 
{\rm d} \xi\, 
\frac{{\bar \Delta}^{2}}
{\{\xi^{2}+(\hbar \varepsilon_{n})^{2}
+{\bar \Delta}^{2}\}^{2}},
\nonumber\\
&=&\frac{2 {\bar \Delta}^{2}}{\beta}
\sum_{\varepsilon_{n}}
\{(\hbar \varepsilon_{n})^{2}
+{\bar \Delta}^{2}\}^{-\frac{3}{2}}
\nonumber\\
&&\phantom{aaaaa}
\times \arctan 
\frac{\Lambda}{\sqrt{(\hbar \varepsilon_{n})^{2}+
{\bar \Delta}^{2}}}. 
\label{f_s}
\end{eqnarray}
Then from Eq. (\ref{i1}) and (\ref{i2}) we obtain 
Eq. (\ref{gg}). 

In a similar way we obtain,  
\begin{eqnarray}
&&\frac{1}{\beta \hbar L}
\sum_{k,\omega_{n}}
{\cal F}(k,i\varepsilon_{n})^{2}
=\frac{I_{2}}{2}, 
\end{eqnarray}
which reduces to Eq. (\ref{ff}). 

Next we examine the behavior of $f_{\rm s}$ 
in two limiting cases, 
$\beta \bar{\Delta} \ll 1$ and 
$\beta \bar{\Delta} \gg 1$. 

\subsection{
The case of $\beta \bar{\Delta} \ll 1$}

In this case, we can simply expand Eq. (\ref{f_s}) 
with respect to $\bar{\Delta}$ and obtain, 
\begin{eqnarray}
f_{\rm s} &\simeq& \frac{1}{\beta}
\sum_{\varepsilon_{n}}\frac{\pi \bar{\Delta}^{2}}
{(\hbar \varepsilon_{n})^{3}}
= \frac{(\beta \bar{\Delta})^{2}}
{4 \pi^{2}}
\sum_{n=1}^{\infty}\frac{1}{(n - \frac{1}{2})^{3}}
\nonumber\\
&=& \frac{(\beta \bar{\Delta})^{2}}
{4 \pi^{2}} \zeta(3,\frac{1}{2}). 
\end{eqnarray}

\subsection{
The case of $\beta \bar{\Delta} \gg 1$}

In this case the following method is 
more useful \cite{Abrikosov}. 
From Eq. (\ref{f_s}), 
we obtain,
\begin{eqnarray}
f_{\rm s} &=& 
\frac{4}{\beta}
\sum_{\varepsilon_{n}}
\int_{0}^{\Lambda}
{\rm d} \xi\,
\frac{\bar{\Delta}^{2}}
{\{\xi^{2}+(\hbar \varepsilon_{n})^{2}
+\bar{\Delta}^{2}\}^{2}},
\nonumber\\
&=& 2 \bar{\Delta}\beta
\int_{0}^{\Lambda} {\rm d} \xi\, 
\frac{\partial}{\partial \bar{\Delta}}
\sum_{\varepsilon_{n}}
\frac{-1}{(2 n-1)^{2} \pi^{2} +
\beta^{2} (\xi^{2}+\bar{\Delta}^{2})},
\nonumber\\
&=& - \bar{\Delta}
\frac{\partial}{\partial \bar{\Delta}}
\int_{0}^{\Lambda}{\rm d} \xi\,
\frac{1}{E}\tanh \frac{\beta E}{2},
\nonumber\\
&=& - \bar{\Delta}
\frac{\partial}{\partial \bar{\Delta}}
\int_{0}^{\Lambda}{\rm d} \xi\,
\frac{1}{E}
\left\{
1 + 2 \sum_{m=1}^{\infty}(-1)^{m}
{\rm e}^{-\beta E m}\right\},
\nonumber\\
&=&
- \bar{\Delta}
\frac{\partial}{\partial \bar{\Delta}}
\biggl[\log\biggl(\frac{\Lambda + 
\sqrt{\Lambda^{2}+\bar{\Delta}^{2}}}
{\bar{\Delta}}
\biggr)
\nonumber\\
&&
+2 \sum_{m=1}^{\infty}
(-1)^{m}
\int_{1}^{\sqrt{\Lambda^{2}+\bar{\Delta}^{2}}
/\bar{\Delta}}{\rm d}\epsilon\,
\frac{1}{\sqrt{\epsilon^{2}-1}}
{\rm e}^{-\beta \bar{\Delta}\epsilon m}
\biggr],
\nonumber\\
&\longrightarrow&
1 + 2 \beta \bar{\Delta}
\sum_{m=1}^{\infty}(-1)^{m}
m K_{1}(\beta \bar{\Delta} m)
\phantom{aa}(\Lambda \longrightarrow \infty),
\nonumber\\
&\simeq&
1 - \sqrt{2 \pi \beta \bar{\Delta}}
{\rm e}^{-\beta \bar{\Delta}},
\label{lowT}
\end{eqnarray}
where 
$E = \sqrt{\xi^{2}+\bar{\Delta}^{2}}$ and 
$\epsilon = E/\bar{\Delta}$. 
$K_{1}(z)$ is the modified Bessel function given by, 
\begin{eqnarray}
K_{1}(z) &=& - \frac{d}{d z} K_{0}(z)
\nonumber\\
&=& - \frac{d}{d z} \int_{1}^{\infty}{\rm d}y\, 
\frac{1}{\sqrt{y^{2}-1}}{\rm e}^{-zy}. 
\end{eqnarray}
In the last line of Eq. (\ref{lowT}), 
we have employed the asymptotic form of $K_{1}(z)$.


\begin{references}
\bibitem{hopping} 
Hopping of electrons between chains 
is assumed to be small but non-zero. 
If it is absent, electrons on each chain become 
independent and chemical potential 
can differ from chain to chain. 
We consider a situation with a finite hopping 
and all the chains are in one equilibrium state as a 
whole.
\end{references}
\end{document}